\documentclass[12pt]{article}

\usepackage[margin=1in,dvips,a4paper,nohead]{geometry}

\makeatletter
\def\eqnarray{%
   \stepcounter{equation}%
   \def\@currentlabel{\p@equation\theequation}%
   \global\@eqnswtrue
   \m@th
   \global\@eqcnt\z@
   \tabskip\@centering
   \let\\\@eqncr
   $$\everycr{}\halign to\displaywidth\bgroup
       \hskip\@centering$\displaystyle\tabskip\z@skip{##}$\@eqnsel
      &\global\@eqcnt\@ne\hfil$\displaystyle{\hbox{}##\hbox{}}$\hfil
      &\global\@eqcnt\tw@ $\displaystyle{##}$\hfil\tabskip\@centering
      &\global\@eqcnt\thr@@ \hb@xt@\z@\bgroup\hss##\egroup
         \tabskip\z@skip
      \cr
}
\makeatother

\usepackage{amsfonts}

\mathchardef\Gamma="100
\def\gothg{\mathfrak g}

\def\Re {\mathop{\mathrm{Re}}}

\title{\bf On Classification of Non-Abelian Toda Systems}

\author{\bf Kh. S. Nirov\\
\small \em Institute for Nuclear Research of the Russian Academy of
Sciences\\[-.3em]
\small \em 60th October Anniversary Prospect 7a, 117312 Moscow,
Russia\\[.3em]
\bf A. V. Razumov\\
\small \em Institute for High Energy Physics\\[-.3em]
\small \em 142281 Protvino, Moscow Region, Russia}

\date{}

\begin{document}

\maketitle

\begin{abstract}
A simple procedure to enumerate all Toda systems associated with complex
classical Lie groups is given.
\end{abstract}

\section{Introduction}

By a Toda system we mean a system of nonlinear partial
differential equations for functions of two real variables or one
complex variable having a special form. A~concrete Toda system is
specified by the choice of a Lie group and by the choice of a
$\mathbb Z$-gradation of the corresponding Lie algebra, see, for
example, \cite{LeSa92, RaSa97a}. Any Toda system is exactly or completely
integrable, and this is actually enough to justify the necessity to
investigate them. As a matter of fact, they arise in many mathematical
and physical problems having as fundamental as application significance.

In this talk we discuss the classification of non-abelian Toda systems
associated with classical Lie groups. Such a classification was
performed earlier with the help of the root decomposition of the
corresponding Lie algebras in papers \cite{RaSa97b, RSZu99}. Here we
use the approach which is not based on root techniques and appeals
only to general properties of simple Lie algebras.

\section{Toda systems}

Let $M$ be either the real manifold $\mathbb R^2$ or the
complex manifold $\mathbb C$. Denote the standard coordinates on
$\mathbb R^2$ by $z^-$ and $z^+$. In the case of the manifold $\mathbb
C$ we denote by $z^-$ the standard complex coordinate $z$ and by $z^+$
its complex conjugate $\bar z$.\footnote{Actually one can assume that
$M$ is an arbitrary two-dimensional real manifold or one-dimensional
complex manifold. Here $z^-$ and $z^+$ are some local coordinates.}

Recall that a Lie algebra $\mathfrak g$ is said to be $\mathbb Z$-graded
if there is given a representation
\[
\mathfrak g = \bigoplus_{m \in \mathbb Z} \mathfrak g_m,
\]
where
\[
[\mathfrak g_m, \mathfrak g_n] \subset \mathfrak g_{m+n}
\]
for all $m, n \in \mathbb Z$. The subspace $\gothg_0$ is a subalgebra
of $\gothg$.

Consider a real or complex matrix Lie group\footnote{The case of a
general, not necessarily matrix, Lie group is considered in the paper
\cite{RaSa94}, and in the book~\cite{RaSa97a}.} $G$ whose Lie algebra
$\mathfrak g$ is endowed with a $\mathbb Z$-gradation. Let for some
positive integer $l$ the subspaces $\mathfrak g_{-m}$ and
$\mathfrak g_{+m}$ for $0 < m < l$ be trivial. Denote by $G_0$
the connected Lie subgroup of $G$ corresponding to the subalgebra
$\gothg_0$. The Toda equations are the matrix equation for a mapping
$\gamma$ from $M$ to $G_0$ which have the following form
\[
\partial_+(\gamma^{-1} \partial_- \gamma) = [c_-, \gamma^{-1}
c_+ \gamma].
\]
Here $c_-$ and $c_+$ are some fixed mappings from $M$ to $\mathfrak g_
{-l}$ and $\mathfrak g_{+l}$ respectively, satisfying the conditions
\[
\partial_+ c_- = 0, \qquad \partial_- c_+ = 0.
\]
When the Lie group $G_0$ is abelian the corresponding Toda system is
said to be abelian, otherwise we deal with a nonabelian Toda system.

There exist the so-called higher grading \cite{Lezn95,
GeSa95, BlFe00, BFRa02} and multi-dimensional \cite{RaSa97c, RaSa97d}
generalisations
of the Toda systems.

\section{$\mathbb Z$-gradations}

It is clear that to classify Toda systems one has to classify
$\mathbb Z$-gradations of Lie algebras. Let us restrict ourselves to
the case of Lie algebras corresponding to classical Lie groups. These
are complex special linear, orthogonal and symplectic groups. All
these groups and the corresponding Lie algebras are simple. This fact
allows one to describe all $\mathbb Z$-gradations. Here the following
facts are used.

Let us have some $\mathbb Z$-gradation of a Lie algebra $\mathfrak g$.
Define a linear operator $D$ acting on an element $x = \sum_{m \in
\mathbb Z} x_m$ as
\[
D x = \sum_{m \in \mathbb Z} m x_m.
\]
The operator $D$ satisfies the relation
\[
D([x, y]) = [Dx, y] + [x, Dy].
\]
Hence, $D$ is a derivation of $\mathfrak g$. For any element of
$x \in \mathfrak g$ the linear operator $\mathrm{ad}(x)$ defined by
\[
\mathrm{ad}(x) y = [x, y]
\]
is a derivation of $\mathfrak g$. Such a derivation is said to be
internal. It is important for our consideration that any derivation of
a semisimple Lie algebra is internal. Therefore, in the case when
$\mathfrak g$ is semisimple, for any $\mathbb Z$-gradation there
exists an element $q$ of $\mathfrak g$ such
that
\[
D x = \mathrm{ad}(q) x = [q, x].
\]
The operator $\mathrm{ad}(q)$, or the element $q$ itself, is called
the grading operator, corresponding to the $\mathbb Z$-gradation under
consideration. If the grading operator exists then the subspaces
$\mathfrak g_m$ can be described as
\[
\mathfrak g_m = \{x \in \mathfrak g \mid [q, x] = mx\}.
\]

Note now that if $q$ is the grading operator generating some
$\mathbb Z$-gradation, then the operator $\mathrm{ad}(q)$ is
diagonalisable. Let us recall the following fact from the theory
of semisimple Lie algebras.

Let $\mathfrak g$ be a complex semisimple Lie algebra,
and $\varphi$ be its linear representation. If for some
$x\in \mathfrak g$ the operator $\mathrm{ad}(x)$ is diagonalisable,
then the linear operator $\varphi(x)$ is also diagonalisable.

In the case of classical Lie algebras we always have a special
representation, it is the defining representation. The above facts say
in this case that any $\mathbb Z$-gradation is generated by the
corresponding grading operator, and that this operator, up to
an automorphism of the Lie algebra under consideration, is a
diagonal matrix.

Now, let us proceed to the description of concrete $\mathbb Z$-gradations
and corresponding Toda systems.

\section{Special linear groups}

We start with the Lie groups $\mathrm{SL}_n(\mathbb C)$ and the
corresponding Lie algebras $\mathfrak{sl}_n(\mathbb C)$.

The grading operator $q$ corresponding to a $\mathbb Z$-gradation of
$\mathfrak{sl}_n(\mathbb C)$ is a diagonalisable matrix. Hence,
up to an internal automorphism any grading operator $q$ has the block
matrix form
\begin{equation}
q = \left( \begin{array}{cccc}
\rho_1 I_{k_1} & 0 & \cdots & 0 \\
0 & \rho_2 I_{k_2} & \cdots & 0 \\
\vdots & \vdots & \ddots & \vdots \\
0 & 0 & \cdots & \rho_p I_{k_p}
\end{array} \right), \label{6}
\end{equation}
where $I_k$ is the $k \times k$ identity matrix, and we assume that $
\Re \rho_1 \ge \Re \rho_2 \ge \ldots \ge \Re \rho_p$. The grading
operator $q$ of the above form belongs to the Lie algebra
$\mathfrak{sl}_n(\mathbb C)$ if $\sum_{a=1}^p k_a = n$ and
\begin{equation}
\sum_{a=1}^p \rho_a k_a = 0. \label{5}
\end{equation}
Represent a general element $x$ of $\mathfrak{sl}_n(\mathbb C)$ in the
block matrix form
\begin{equation}
x = \left( \begin{array}{ccc}
x_{11} & \cdots & x_{1p} \\
\vdots & \ddots & \vdots \\
x_{p1} & \cdots & x_{pp}
\end{array} \right), \label{8}
\end{equation}
where $x_{ab}$ is a $k_a \times k_b$ matrix. It is clear that
\[
[q, x]_{ab} = (\rho_a - \rho_b) x_{ab}.
\]
Hence, since $q$ generates a $\mathbb Z$-gradation we should have $m_a
= \rho_a - \rho_{a+1} \in \mathbb Z$. It follows that all numbers
$\rho_a$ have the same imaginary part, and in reality the equality
(\ref{5}) implies that they are real and we will assume that
$\rho_1 > \rho_2 > \ldots > \rho_p$. The numbers $\rho_a$ can be expressed
via the integers $k_a$ and $m_a$:
\begin{equation}
\rho_a = \frac{1}{n} \left( - \sum_{b=1}^{a-1} m_b \sum_{c=1}^b k_c +
\sum_{b=a}^{p-1} m_b \sum_{c=b+1}^p k_c \right). \label{7}
\end{equation}
Thus a $\mathbb Z$-gradation of the Lie algebra
$\mathfrak{sl}_n(\mathbb C)$ is uniquely specified by the choice of
$p$ integers $k_a$ such that $\sum_{a=1}^p k_a = n$ and by the choice
of $p-1$ positive integers $m_a$. The grading operator $q$ has the form
(\ref{6}), where the numbers $\rho_a$ are given by the relation (\ref{7}).

The grading structure of the Lie algebras $\mathfrak{sl}_n(\mathbb C)$ can
be depicted by the following scheme
\[
\left( \begin{array}{c|c|c|c|c}
0 & m_1 & m_1 + m_2 & \cdots & \displaystyle \sum_{a=1}^{p-1}m_a \\
\hline
-m_1 & 0 & m_2 & \cdots & \displaystyle \sum_{a=2}^{p-1}m_a \\
\hline
-(m_1 + m_2) & -m_2 & 0 & \cdots & \displaystyle \sum_{a=3}^{p-1}m_a
\\
\hline
\vdots & \vdots & \vdots & \ddots & \vdots \\
\hline
-\displaystyle \sum_{a=1}^{p-1}m_a & -\displaystyle
\sum_{a=2}^{p-1}m_a & -\displaystyle \sum_{a=3}^{p-1}m_a & \cdots & 0
\end{array} \right).
\]
Here the numbers in the boxes correspond to the grading indices of the
corresponding blocks in the block matrix representation (\ref{8}) of a
general element of $\mathfrak{sl}_n(\mathbb C)$. Note, in particular, that
the subalgebra $\gothg_0$ is formed by all block diagonal matrices.
The group $G_0$ is also formed by block diagonal matrices and is isomorphic
to $\mathrm{GL}_{k_1}(\mathbb C) \times \cdots \times
\mathrm{GL}_{k_p}(\mathbb C)$.

Consider now Toda systems associated with the Lie group
$\mathrm{SL}_n(\mathbb C)$. Actually, it is more convenient to deal with
the Lie group $\mathrm{GL}_n(\mathbb C)$. Any grading operator for the Lie
algebra $\mathfrak{sl}_n(\mathbb C)$ can be considered as a grading
operator for the Lie algebra $\mathfrak{gl}_n(\mathbb C)$.

It can be easily understood that to exhaust all Toda systems it suffices to
consider only gradations with all numbers $m_a$ equal to $1$ \cite{RSZu99}.
In this case the mappings $c_-$ and $c_+$ should take values in the
subspaces $\mathfrak g_{-1}$ and $\mathfrak g_{+1}$ respectively. The
general forms of such elements are
\begin{equation}
c_- = \left( \begin{array}{ccccc}
0 & 0 & \cdots & 0 & 0 \\
C_{-1} & 0 & \cdots & 0 & 0 \\
\vdots & \vdots & \ddots & \vdots & \vdots \\
0 & 0 & \cdots & 0 & 0 \\
0 & 0 & \cdots & C_{-(p-1)} & 0
\end{array} \right), \qquad
c_+ = \left( \begin{array}{ccccc}
0 & C_{+1} & \cdots & 0 & 0 \\
0 & 0 & \cdots & 0 & 0 \\
\vdots & \vdots & \ddots & \vdots & \vdots \\
0 & 0 & \cdots & 0 & C_{+(p-1)} \\
0 & 0 & \cdots & 0 & 0
\end{array} \right), \label{11}
\end{equation}
where for each $a = 1, \ldots, p-1$ the mapping $C_{-a}$ takes
values in the space of $k_{a+1} \times k_a$ complex matrices, and the
mapping $C_{+a}$ takes values in the space of $k_a \times k_{a+1}$
complex matrices. Besides, these mappings must satisfy the relations
\[
\partial_+ C_{-a} = 0, \qquad \partial_- C_{+a} = 0.
\]

\noindent Parametrise the mapping $\gamma$ as
\begin{equation}
\gamma = \left( \begin{array}{cccc}
\Gamma_1 & 0 & \cdots & 0 \\
0 & \Gamma_2 & \cdots & 0 \\
\vdots & \vdots & \ddots & \vdots \\
0 & 0 & \cdots & \Gamma_p
\end{array} \right), \label{12}
\end{equation}
where the mappings $\Gamma_a$ take values in the Lie groups
$\mathrm{GL}_{\, k_a}(\mathbb C)$. In this parametrisation Toda
equations take the form
\[ \arraycolsep 0pt
\begin{array}{rclcl}
\partial_+ \left( \Gamma_1^{-1} \, \partial_- \Gamma_1^{} \right) & {}
= {} & {} -\Gamma_1^{-1} \, C_{+1}^{} \, \Gamma_2^{} \, C_{-1}^{}, & &
\\[.5em]
\partial_+ \left( \Gamma_a^{-1} \, \partial_- \Gamma_a^{} \right) & {}
= {} & {} -\Gamma_a^{-1} \, C_{+a}^{} \, \Gamma_{a+1}^{} \, C_{-a}^{} &{}+
{}& C_{-(a-1)}^{} \, \Gamma_{a-1}^{-1} \, C_{+(a-1)}^{} \, \Gamma_a^{},
\quad 1 < a < p, \\[.5em]
\partial_+ \left( \Gamma_p^{-1} \, \partial_- \Gamma_p^{} \right) & {}
= {} & & & C_{-(p-1)}^{} \, \Gamma_{p-1}^{-1} \, C_{+(p-1)}^{} \,
\Gamma_p^{}.
\end{array}
\]
The simplest case is when one chooses $k_a = k$ and $C_{-a} = C_{+a} =
I_k$:
\[ \arraycolsep 0pt
\begin{array}{rclcl}
\partial_+ \left( \Gamma_1^{-1} \, \partial_- \Gamma_1^{} \right) & {}
= {} & {} -\Gamma_1^{-1} \, \Gamma_2^{}, && \\[.5em]
\partial_+ \left( \Gamma_a^{-1} \, \partial_- \Gamma_a^{} \right) & {}
= {} & {} - \Gamma_a^{-1} \, \Gamma_{a+1}^{} & {} + {} & \Gamma_{a-1}^{-1}
\, \Gamma_a^{}, \quad 1 < a < p, \\[.5em]
\partial_+ \left( \Gamma_p^{-1} \, \partial_- \Gamma_{p}^{} \right) & {}
= {} & & & \Gamma_{p-1}^{-1} \, \Gamma_p^{}.
\end{array}
\]

\section{Orthogonal Lie groups}

It is convenient for our purposes to define the complex orthogonal group
$\mathrm O_n(\mathbb C)$ as the Lie subgroup of $\mathrm{GL}_n(\mathbb C)$
formed by the elements $a \in \mathrm{GL}_n(\mathbb C)$ satisfying the
condition
\begin{equation}
a^t J_n a = J_n, \label{9}
\end{equation}
where $J_n$ is the skew-diagonal $n \times n$ unit matrix. The Lie algebra
$\mathfrak o_n(\mathbb C)$ of the Lie group $\mathrm O_n(\mathbb C)$
consists of $n \times n$ complex matrices $x$ satisfying the condition
\begin{equation}
x^t J_n + J_n x = 0. \label{10}
\end{equation}
For a $k_1 \times k_2$ matrix $a$ we will denote
\[
a^T = J_{k_2} a^t J_{k_1}.
\]
Actually, $a^T$ is the transpose of $a$ with respect to the skew diagonal.
The conditions (\ref{9}) and (\ref{10}) can be written now as
$a^T = a^{-1}$ and $x^T = -x$, respectively.

The Lie algebras $\mathfrak o_n(\mathbb C)$ are simple and it is clear
that any $\mathbb Z$-gradation of $\mathfrak o_n(\mathbb C)$ is generated
by the corresponding grading operator, which has the form (\ref{6}) and
belongs to $\mathfrak o_n(\mathbb C)$. Hence, a $\mathbb Z$-gradation of
the Lie algebra $\mathfrak o_n(\mathbb C)$ is uniquely specified by the
choice of $p$ integers $k_a$ such that $\sum_{a=1}^p k_a = n$, $k_a =
k_{p-a+1}$, and by the choice of $p-1$ positive integers $m_a$ such
that $m_a = m_{p-a}$.

Consider now the Toda equations associated with the Lie groups
$\mathrm O_n(\mathbb C)$. The general form of the mappings $c_-$ and $c_+$
is again given by (\ref{11}), but here the mappings
$C_{\pm a}$ should obey the relations
\begin{equation}
C_{-a}^T = - C_{-(p-a)}, \qquad C_{+a}^T = -C_{+(p-a)}. \label{13}
\end{equation}
The Lie group $G_0$ in the case $p=2s-1$ is isomorphic to
$\mathrm{GL}_{k_1}(\mathbb C) \times \cdots \times
\mathrm{GL}_{k_{s-1}}(\mathbb C) \times \mathrm{SO}_{k_s}(\mathbb C)$
while in the case $p = 2s$ it is isomorphic to
$\mathrm{GL}_{k_1}(\mathbb C) \times \cdots \times
\mathrm{GL}_{k_s}(\mathbb C)$. We can use the same
parametrisation (\ref{12}) for $\gamma$ as for the case of the Lie group
$\mathrm{GL}_n(\mathbb C)$. Here one has
\begin{equation}
\Gamma_a^T = \Gamma_{p-a+1}^{-1}. \label{14}
\end{equation}
The Toda equations have the same form as for the case of the Lie groups $
\mathrm{GL}_n(\mathbb C)$. One only has to take into account the relations
(\ref{13}) and (\ref{14}). In the case of $p = 2s - 1$ we have $s$
independent mappings $\Gamma_a$ and the equations for them are
\begin{eqnarray*}
\partial_+ \left( \Gamma_1^{-1} \, \partial_- \Gamma_1^{} \right) &=& {} -
\Gamma_1^{-1} \, C_{+1}^{} \, \Gamma_2^{} \, C_{-1}, \\
\partial_+ \left( \Gamma_a^{-1} \, \partial_- \Gamma_a^{} \right) &=& {} -
\Gamma_a^{-1} \, C_{+a}^{} \, \Gamma_{a+1}^{} \, C_{-a} + C_{-(a-1)} \,
\Gamma_{a-1}^{-1} \, C_{+(a-1)} \, \Gamma_a^{}, \quad 1 < a < s, \\
\partial_+ \left( \Gamma_s^{-1} \, \partial_- \Gamma_s^{} \right) &=& {} -
\Gamma_s^{T} \, C_{+(s-1)}^T \, \Gamma_{s-1}^{-1T} \, C_{-(s-1)}^T +
C_{-(s-1)}^{} \, \Gamma_{s-1}^{-1} \, C_{+(s-1)}^{} \, \Gamma_s^{}.
\end{eqnarray*}
Stress on that in this case $\Gamma_s^T = \Gamma_s^{-1}$. In the case
$p = 2s$ we have the equations
\begin{eqnarray*}
\partial_+ \left( \Gamma_1^{-1} \, \partial_- \Gamma_1{} \right) &=& {} -
\Gamma_1^{-1} \, C_{+1}^{} \, \Gamma_2^{} C_{-1}^{}, \\
\partial_+ \left( \Gamma_a^{-1} \, \partial_- \Gamma_a^{} \right) &=& {} -
\Gamma_a^{-1} \, C_{+a}^{} \, \Gamma_{a+1}^{} \, C_{-a} + C_{-(a-1)}^{} \,
\Gamma_{a-1}^{-1} \, C_{+(a-1)}^{} \, \Gamma_a^{}, \quad 1 < a < s, \\
\partial_+ \left( \Gamma_s^{-1} \, \partial_- \Gamma_s^{} \right) &=& {} -
\Gamma_s^{-1} \, C_{+s}^{} \, \Gamma_s^{-1T} \, C_{-s}^{} + C_{-(s-1)} \,
\Gamma_{s-1}^{-1} \, C_{+(s-1)}^{} \, \Gamma_s^{},
\end{eqnarray*}
where the mappings $C_{\pm s}$ satisfy the relations
\begin{equation}
C_{-s}^T = - C_{-s}^{}, \qquad C_{+s}^T = - C_{+s}^{}. \label{15}
\end{equation}

\section{Symplectic Lie groups}

Define the Lie group $\mathrm{Sp}_{2n}(\mathbb C)$ as the Lie subgroup
of the Lie group $\mathrm{GL}_{2n}(\mathbb C)$ formed by the elements
$a \in \mathrm{GL}_{2n}(\mathbb C)$ which satisfy the condition
\[
a^t K_{2n} a = K_{2n},
\]
where
\[
K_{2n} = \left( \begin{array}{rr}
0 & J_n \\
-J_n & 0
\end{array} \right).
\]
Then the Lie algebra $\mathfrak{sp}_{2n}(\mathbb C)$ of
$\mathrm{Sp}_{2n}(\mathbb C)$ is formed by all $2n \times 2n$
complex matrices $x$ satisfying the condition
\[
x^t K_{2n} + K_{2n} x = 0.
\]
The Lie algebras $\mathfrak{sp}_{2n}(\mathbb C)$ are simple and any
$\mathbb Z$-gradation of $\mathfrak{sp}_{2n}(\mathbb C)$ is generated
by the grading operator, which has the form (\ref{6}) and belongs to
$\mathfrak{sp}_{2n}(\mathbb C)$. One can get convinced that a
$\mathbb Z$-gradation of the Lie algebra $\mathfrak{sp}_{2n}(\mathbb C)$
is uniquely specified by the same data as for a $\mathbb Z$-gradation of
the Lie algebra $\mathfrak o_{2n}(\mathbb C)$.

Consider now the corresponding Toda equations. The mappings $c_-$ and $c_+$
have the forms (\ref{11}) with the mappings $C_{\pm a}$ satisfying the
relations
\begin{eqnarray*}
&& C_{-a}^T = -C_{-(p-a)}^{}, \qquad C_{+a}^T = -C_{+(p-a)}^{}, \qquad a
\ne s-1,s, \\
&& J_{k_{s-1}} \, C_{-(s-1)}^t \, K_{k_s} = - C_{-s}^{}, \qquad K_{k_s} \,
C_{+(s-1)}^t \, J_{k_{s-1}} = - C_{+s}^{}
\end{eqnarray*}
when $p = 2s-1$, and
\begin{eqnarray*}
&& C_{-a}^T = -C_{-(p-a)}^{}, \qquad C_{+a}^T = -C_{+(p-a)}^{}, \qquad a
\ne s, \\
&& C_{-s}^T = C_{-s}^{}, \qquad C_{+s}^T = C_{+s}^{},
\end{eqnarray*}
when $p = 2s$.

The Lie group $G_0$ in the case of $p=2s-1$ is isomorphic to
$\mathrm{GL}_{k_1}(\mathbb C) \times \cdots \times
\mathrm{GL}_{k_{s-1}}(\mathbb C) \times \mathrm{Sp}_{k_s}(\mathbb C)$,
while in the case of $p = 2s$ it is isomorphic to
$\mathrm{GL}_{k_1}(\mathbb C) \times \cdots \times
\mathrm{GL}_{k_s}(\mathbb C)$.
We can use the parametrisation (\ref{12}) for the mapping $\gamma$.
Here, in the case of $p = 2s - 1$, one has
\[
\Gamma_a^T = \Gamma_{p-a+1}^{-1}, \quad a \ne s+1, \qquad
\Gamma_s^t \, K_{k_s} \, \Gamma_s^{} = K_{k_s},
\]
whereas in the case of $p = 2s$
\[
\Gamma_a^T = \Gamma_{p-a+1}^{-1}
\]
for any $a$. The independent Toda equations in the case of $p = 2s-1$ have
the form
\begin{eqnarray*}
\partial_+ \left( \Gamma_1^{-1} \, \partial_- \Gamma_1^{} \right) &=& {} -
\Gamma_1^{-1} \, C_{+1} \, \Gamma_2^{} C_{-1}^{}, \\
\partial_+ \left( \Gamma_a^{-1} \, \partial_- \Gamma_a^{} \right) &=& {} -
\Gamma_a^{-1} \, C_{+a}^{} \, \Gamma_{a+1}^{} \, C_{-a}^{} + C_{-(a-1)}^{}
\Gamma_{a-1}^{-1} \, C_{+(a-1)}^{} \, \Gamma_a^{}, \quad 1 < a < s, \\
\partial_+ \left( \Gamma_s^{-1} \, \partial_- \Gamma_s^{} \right) &=&
\Gamma_s^{-1} \, K_{k_s} \, C_{+(s-1)}^t \, \Gamma_{s-1}^{-1t} \,
C_{-(s-1)}^t \, K_{k_s} + C_{-(s-1)}^{} \, \Gamma_{s-1}^{-1} \,
C_{+(s-1)}^{} \, \Gamma_s^{}.
\end{eqnarray*}
In the case of $p = 2s$ one has the same equations as for the Lie groups
$\mathrm O_{2n}(\mathbb C)$, but with the mappings $C_{\pm s}$ satisfying
instead of (\ref{15}) the relations
\[
C_{-s}^T = C_{-s}^{}, \qquad C_{+s}^T = C_{+s}^{}.
\]

\section{Simplest example}

A first non-abelian Toda system which was integrated explicitly was
a system associated with the Lie group $\mathrm{O}_5(\mathbb C)$. We
start this section with the description of this system along lines
used in early papers \cite{LeSa83, GeSa92, Bila94}. The Lie algebra $
\mathfrak{o}_5(\mathbb C)$ of the Lie group $\mathrm{O}_5(\mathbb C)$
is of type $B_2$. Let $h_1$, $h_2$ be Cartan generators, and
$x_{\pm 1}$, $x_{\pm 2}$ be the Chevalley generators corresponding
to the simple roots $\alpha_1$, $\alpha_2$. Consider the
$\mathbb Z$-gradation generated by the grading operator
$q = 2 h_1 + h_2$. The grading subspaces have the forms
\begin{eqnarray*}
\gothg_{-1} &=& \mathbb C \, \gothg^{- \alpha_1} \oplus \mathbb C \,
\gothg^{- \alpha_1 - \alpha_2} \oplus \mathbb C \, \gothg^{- 2 \alpha_
1 - \alpha_2} \\
\gothg_0 &=& \mathbb C \, \gothg^{- \alpha_2} \oplus \mathfrak h
\oplus \mathbb C \, \gothg^{\alpha_2} \\
\gothg_{+1} &=& \mathbb C \, \gothg^{\alpha_1} \oplus \mathbb C \,
\gothg^{\alpha_1 + \alpha_2} \oplus \mathbb C \, \gothg^{2 \alpha_1 +
\alpha_2}.
\end{eqnarray*}
Choose the fixed mappings $c_-$ and $c_+$ as
\[
c_+ = [x_{+1}, x_{+2}], \qquad c_- = [x_{-2}, x_{-1}],
\]
and parametrise the mapping $\gamma$ as
\[
\gamma = \exp(a_+ x_{+2}) \exp(a_- x_{-2}) \exp(a_1 h_1 + a_2 h_2)
\]
Then the Toda equations take the form
\begin{eqnarray}
\partial_+ \partial_- a_1 &=& {} - 2 \mathrm e^{-a_1} (1 + 2 a_- a_+),
\label{1} \\
\partial_+(\partial_- a_1 - \partial_- a_2 - a_- \partial_- a_+) &=&
{} - \mathrm e^{-a_2} (1 + 2 a_- a_+), \label{2} \\
\partial_+(\mathrm e^{a_1 - 2 a_2} \partial_- a_+) &=& 2
\mathrm e^{- 2 a_2} a_+, \label{3} \\
\partial_+[\mathrm e^{- a_1 + 2 a_2} (\partial_- a_- - a_-^2
\partial_- a_+)] &=& 2 \mathrm e^{-2a_1 + 2a_2} a_- (1 + a_- a_+).
\label{4}
\end{eqnarray}

Now, let us show how the system described above enters into our
classification of non-abelian Toda systems. First of all notice that
the Lie group $\mathrm{O}_5(\mathbb C)$ is locally isomorphic to
the Lie group $\mathrm{Sp}_4(\mathbb C)$. Moreover, the formulation of
the system was based only on local properties of the Lie group $
\mathrm{O}_5(\mathbb C)$. Therefore, we will come to the same system
starting from the Lie group $\mathrm{Sp}_4(\mathbb C)$. This also
allows one to include the system under consideration in a series
of non-abelian Toda systems associated with the Lie groups
$\mathrm{Sp}_{2n}(\mathbb C)$ which have an extremely simple form.

Endow the Lie algebra $\mathfrak{sp}_{2n}(\mathbb C)$ with
a $\mathbb Z$-gradation generated by the grading operator
\[
q = \frac{1}{2} \left( \begin{array}{cc}
I_n & 0 \\
0 & -I_n
\end{array} \right).
\]
The parametrisation of the mapping $\gamma$ corresponding with this
$\mathbb Z$-gradation has the form
\[
\gamma = \left( \begin{array}{cc}
\Gamma & 0 \\
0 & (\Gamma^T)^{-1}
\end{array} \right).
\]
Choose the mappings $c_-$ and $c_+$ as
\[
c_- = \left( \begin{array}{cc}
0 & 0 \\
I_n & 0
\end{array} \right), \qquad
c_+ = \left( \begin{array}{cc}
0 & I_n \\
0 & 0
\end{array} \right).
\]
The Toda equations have in our case the form
\begin{equation}
\partial_+(\Gamma^{-1} \partial_- \Gamma) = - (\Gamma^T
\Gamma)^{-1}. \label{16}
\end{equation}

In the case of $n=2$, if one parametrises $\Gamma$ as
\[
\Gamma = \left( \begin{array}{cc}
\mathrm e^{a_2}(1 + a_- a_+) & \mathrm e^{a_1 - a_2} a_+ \\[.5em]
\mathrm e^{a_2} a_- & \mathrm e^{a_1 - a_2}
\end{array}\right),
\]
then the Toda equations (\ref{16}) take the form (\ref{1})--(\ref{4}).
Certainly, the equation (\ref{16}) is more attractive than the equations
(\ref{1})--(\ref{4}).

In the case of $n=1$ the mapping $\Gamma$ is just a function, and we have
the equation
\[
\partial_+(\Gamma^{-1} \partial_- \Gamma) = - \Gamma^{-2}.
\]
Introducing a function $F$ via the relation
\[
\Gamma = \mathrm e^{F}
\]
one comes to the famous Liouville equation
\[
\partial_+ \partial_- F = {} -\mathrm e^{-2F}.
\]
Therefore, it is quite natural for general $n$ to call the equation
\[
\partial_+(\Gamma^{-1} \partial_- \Gamma) = - (\Gamma^T
\Gamma)^{-1}
\]
the non-abelian Liouville equation.

\section{Conclusions}

We have shown that the classification of the non-abelian Toda systems
associated with complex classical Lie groups can be performed using only
some general properties of the semisimple Lie algebras. The arising block
matrix structure appears to be very convenient. For example, it allows one
to find explicit forms for $W$-algebras corresponding to non-abelian Toda
systems \cite{BTVr91, NiRa02a, NiRa02b}.

The work of A.V.R. was supported in part by the Russian Foundation
for Basic Research under grant \#01--01--00201 and the INTAS under
grant \#00--00561.

\end{document}